# Crack initiation of printed lines predicted with digital image correlation

C. Katsarelis[1], O. Glushko[2], C.Tonkin[3], M.S. Kennedy[1], M. J. Cordill[2*]

[1]Department of Materials Science & Engineering, Clemson University, Clemson, SC 29634 USA
[2]Erich Schmid Institute of Materials Science, Austrian Academy of Sciences, and Department of Material Physics, Montanuniversität Leoben, Leoben, Austria
[3]Clemson Sonoco Institute for Packaging Design and Graphics, Clemson University, Clemson, SC 29634 USA
*corresponding author: megan.cordill@oeaw.ac.at

**Abstract**

Printing of metallic films has been preferred for roll-to-roll processes over vacuum technologies due to faster processing times and lower processing costs. Films can be produced by depositing inks containing suspended metallic particles within a solvent and then heating the films to both remove the solvent and sinter the particles. The resulting printed structure, electrical and mechanical behavior of the printed films has been studied to better understand their electro-mechanical response to loading and eventual brittle fracture. This study evaluated the electro-mechanical behavior of 1.25 µm printed Ag films using in-situ resistance and in-situ imaging methods. Digital image correlation was utilized with confocal laser scanning microscope images to better visualize crack initiation during tensile straining. This technique showed that cracks initiated earlier in the thicker areas of the film (crests) than in lower areas (troughs) because of a higher density of printing defects and the increased thickness.

**Introduction**

Electronic devices continue to be increasingly integrated into individuals' everyday life, with newly identified applications beginning to change design requirements. New requirements for devices have emerged including the need to reduce the relative electronic device weight and the ability of some devices to undergo bending and stretching while maintaining their electronic functionality. Manufacturing methods for flexible electronics, which are more bendable and/or stretchable than traditional electronic devices, are being developed [1,2] to optimize both the manufacturing throughput and device performance. Each of these flexible electronic devices is composed of a substrate, such as a polyimide or steel sheet that is coated with metallic and or ceramic features. Roll-to-roll (R2R) manufacturing involves the continuous motion of the substrate through a series of rolls and stations [1–5]. Printing of metallic films has been preferred over vacuum-based deposition methods for R2R processes due to relatively faster processing times and lower costs.

To optimize R2R manufacturing processes further, researchers are exploring different film deposition methods including digital inkjet, gravure, and flexographic printing [6–8]. All these methods use an ink with suspended metallic particles and once the ink is deposited, a heating step is applied to both remove the remaining solvent and sinter the metallic particles. During digital inkjet printing a nozzle deposits individual droplets of ink into an array to create a metallic feature [9]. This is in contrast to both gravure and flexographic printing, where the ink is initially transferred to a plate or cylinder prior to the pattern being transferred to the device substrate [9]. In gravure printing, a cylinder is coated with the ink until the recesses are filled and then the excess ink is removed prior to ink being transferred from the recesses onto the substrate [9]. A contact method of transferring is used in the flexographic method but instead of the ink filling recesses in the cylinder prior to deposition, the design is on a soft polymer plate with a raised image of the design wrapped around a cylinder [9]. The ink is then transferred to the raised image and rolled onto the substrate.

The specific parameters of a heating process that removes the remaining solvent and sinters the metallic particles vary for each ink blend depending on the specific components. After heating, the resulting structure of the printed metallic films has been shown to contain pores and relatively high surface roughnesses compared to vacuum deposited films on the same substrates. Studies of their mechanical behavior without a post deposition process suggests that these films behave as brittle materials [6,10,11]. This brittle behavior has been linked to the film porosity where the pores act as stress concentrators and facilitate crack initiation and propagation [6]. However, additional work is needed to evaluate when mechanical damage initiates to provide insight into which factors influence the crack initiation process.

This study is focused on exploring a new approach to tracking mechanical damage in printed films and correlating the damage to the electrical behavior. The use of digital image correlation (DIC) with confocal laser scanning microscope (CLSM) laser intensity images will be demonstrated and how the local mechanical behavior correlates to the global electrical behavior. Printed films have a high surface roughness that creates a natural pattern necessary for DIC techniques. The DIC technique allows for local strain to be calculated by comparing CLSM laser intensity images taken during straining. The local strain measurements are then used to identify areas of high strain that then lead to fracture.

**Experimental Methods**

A flexographic process was used to deposit 1.25 μm thick Ag lines that were 3 mm wide onto a DuPont Teijin Heat Stable substrate (OMET Varylex 530 system with PChem PFI-722 Ag nanoparticle ink). Further details of the printing process were described previously [12] and there was a two year window between film deposition and testing. During this time, the films were not stored in a manner designed to protect them from ultraviolet light.

The normalized resistance of the Ag lines were monitored during monotonic tensile testing and compared to the predicted normalized resistance ($R/R_0$). The relationship between engineering strain and normalized resistance can be approximated by [10,13]:

$$\frac{R}{R_0} = \left(\frac{l}{l_0}\right)^2 \equiv (1 + \varepsilon)^2 \tag{1}$$

where $l$ is the gage length during straining ($l = l_0 + \Delta l$) and it is assumed that the volume of the sample is held constant when there is no phase change ($l \times A = l_0 \times A_0$).

Uniaxial tensile tests were performed on a MTS Tytron 250 universal tensile tester in displacement-controlled mode up to the maximum strain of 30% at a loading rate of $3.3 \times 10^{-4}$ s$^{-1}$. The electric resistance was measured in-situ during straining in four-point probe geometry with the contacts incorporated directly into the grips. More information on the set up has been previously published [10]. Post-straining scanning electron microscopy (SEM) imaging was performed in a Zeiss LEO 1525 workstation.

In-situ tensile straining with CLSM was performed utilizing an Anton Paar TS600 straining stage in conjunction with an Olympus LEXT 4100 OLS for surface imaging. The configuration consisted of the printed lines being strained at a rate of 10 μm/s initially using small straining steps (0.5% and 1% engineering strains) followed by 2% straining steps up to a maximum engineering strain of 12%. At each predetermined step, a 4 min pause was necessary to image the same surface area at three different magnifications. Two types of images were captured with the CLSM technique that tracked either the laser intensity or sample height. For DIC, only the laser intensity images were utilized and the 3-dimensional DIC software GOM Correlate [14] was employed to evaluate the strains parallel to the straining direction. The local surface strains were determined by comparing each image with the initial unstrained image in the straining sequence with the GOM Correlate software.

**Results and Discussion**

A typical result of the relationship between normalized resistance and engineering strain for the printed Ag lines is shown in Figure 1a. During tensile loading of the Ag lines, the normalized resistance initially aligns with the theoretical prediction (Eqn. 1) before deviating to a higher normalized resistance at approximately 3% engineering strain. This deviation to a higher normalized resistance correlates to the printed line reaching a critical cracking strain for brittle materials. The normalized resistance then continued to increase during continued loading, and then, upon unloading, the normalized resistance began to recover. This recovery is due to a combination of elastic recovery and crack re-bridging, when opposing edges of cracks reconnect to some degree [15]. The cracking of the printed Ag films after 30% engineering strain can be seen in as SEM of the surface (Figure 1b). These cracks formed during loading of the films and were not observed in the as-deposited films. The propagation of the cracks is assisted by the pores and pre-existing print defects in the film.

What cannot be observed in Figure 1b is the role that the flexographic printed wavy surface and the initial defects have on the crack initiation and evolution. The printed surface has waves, which range in height (amplitude) between 400 and 800 nm, as measured from CLSM height images (Figure 2a,b). At this magnification it is difficult to evaluate the crack density using the height images, but cracks are visible in the laser intensity image (Figure 2c). A proof-of-concept of the applicability of DIC strain measurement on the flexographic printed lines is shown in Figure 2d. The CLSM laser images (Figure 2c) provide better and more varied contrast for correlating the images than optical or SEM micrographs. From the DIC (Figure 2d), it can be observed that the wavy surface leads to a two different crack densities (qualitatively shown) due to the large thickness variation between the wave crests and troughs. After only 12% engineering strain, the difference can be better viewed the DIC calculated strain image (Figure 2d), which is the same area shown in Figure 2a,c. Comparing the height image and profiles, which are rotated 90 degrees in Figure 2b from Figure 2a, it can be clearly seen that there is a higher density of regions of high strain or cracks (red streaks) in the DIC image that correspond to the wave crests compared the troughs. Thicker films fracture before similar thinner films and thicker films have lower average crack densities (larger average crack spacing) [16–20]. A similar result was observed in the wavy lines by the different crack densities. It should be noted that the larger print defects were found on the wave crests compared to the troughs (Figure 2a) and also influenced crack initiation in the printed lines.

Closer inspection of the CLSM images taken at a higher magnification of a wave crest provide more details about crack initiation and propagation in flexographic printed films. At an engineering strain of 4%, it is difficult to observe cracks in the CLSM laser image (Figure 3a) even though the in-situ resistance measurements indicate fracture of the film (Figure 1). Recall that the resistance measurements are a global measurement while the in-situ CLSM images are a local measurement, concentrated only on a small portion of the entire sample. Application of the DIC local strain measurements provides evidence of crack initiation. In Figure 3b, red streaks, which have an increased local strain, indicate crack initiation near the large defects. The diagonal stripes in Figure 3b are due to the movement of laser during imaging and are artifacts from the DIC analysis that arise only at low applied strains and do not influence the interpreted results. At a higher level of strain (Figure 3c,d) cracks are now more visible in the CLSM laser image (arrows) and correlate well with what is calculated with DIC. Again, more cracks (areas of high strain) are found near the pre-existing defects of the crest. Finally at the maximum applied strain (Figure 3e,f) the crack density has increased and more cracks are observed in both CLSM image and the DIC local strain measurement. However, not only are areas of high strain are found (red streaks) but also more white areas, which correspond to larger cracks on the surface in the CLSM image. The white areas are a result of an extreme contrast difference from the initial image and directly correspond to visible cracks at high strains in the CLSM laser images.

Initially, only DIC can be used to evaluate the crack density (4-6% engineering strain) to directly observe cracks of the rough surface. However, the roughness does make DIC analysis possible to examine the early stages of crack growth in printed films (Figure 4). At about 8% engineering strain, both CLSM and DIC images can be used to determine the crack density and are comparable to one another. The measured crack densities deviate at larger strains, most likely due to the fact that two methods have two different resolutions. Cracks viewed in CLSM images can only be evaluated by eye (what can be seen in the image), while the DIC measurements look at the very local strains on the microscale (pixel resolution). It is interesting that the saturation in the crack density has not been reached at 12% strain and that the shape of the trend is a gradual increase, rather than immediate increase in density commonly observed in sputtered or evaporated films [18,21]. The two different crack density measurements, therefore, give a maximum (DIC) and minimum (CLSM) threshold for crack density until, most likely, saturation is reached at much higher strains. For the current in-situ straining set-up with CLSM, crack saturation could not be achieved since it most likely occurs at engineering strains around or even higher than 30% (Figure 1).

**Summary**

The use of DIC as a characterization method of printed lines added insight into crack initiation and propagation within printed Ag lines. Surface images were collected using the CLSM technique that provided clear, laser intensity images of printed surfaces with relatively large surface roughnesses. The combination of the CLSM laser images and DIC made it possible to more thoroughly examine crack initiation at the local scale in the printed lines. Using DIC, the crack densities were more easily evaluated and showed that cracks initiated earlier in the thicker areas of the film (crests) than in lower areas (troughs) because of a higher density of printing defects and the increased thickness. This insight will be used to alter processing routes to decrease the surface wave amplitudes and printing defects in order to reduce initiation sites for cracks and therefore, indirectly, reduce electrical degradation resulting from applied strain. These results lead the authors to suggest that this technique is valuable when studying the fracture of materials with relatively high surface roughnesses where it is difficult to observe fracture processes.

**Acknowledgements**

The authors gratefully acknowledge support from the Austrian Marshall Plan Scholarship Program for travel between institutions and the opportunity to collaborate. O.G. acknowledges the support from Austrian Science Fund, project P27432-N20. Printed lines in this study were deposited by Mr. S. Rickard from the Sonoco Institute in 2014 under the guidance of Dr. Charles Tonkin. The confocal laser scanning microscopy images were made possible through a New Frontiers Research Infrastructure grant by the Austrian Academy of Sciences (NRFI 2015/02).

**References**

1. National Research Council, *Flexible Electronics for Security, Manufacturing, and Growth in the United States: Summary of a Symposium.* (The National Academies Press., Washington, DC, 2013).
2. W. S. Wong and A. Salleo, *Flexible Electronics: Materials and Applications* (Springer Science and Business Media, 2009).
3. S. H. Ahn and L. J. Guo, Adv. Mater. **20**, 2044 (2008).
4. F. C. Krebs, J. Fyenbo, and M. Jørgensen, J. Mater. Chem. **20**, 8994 (2010).
5. P. Maury, D. Turkenburg, N. Stroeks, P. Giesen, I. Barbu, E. Meinders, A. Van Bremen, N. Iosad, R. Van der Werf, and H. Onvlee, Microelectron. Eng. **88**, 2052 (2011).
6. B.-J. Kim, T. Haas, A. Friederich, J.-H. Lee, D.-H. Nam, J. R. Binder, W. Bauer, I.-S. Choi, Y.-C. Joo, P. a. Gruber, and O. Kraft, Nanotechnology **25**, 125706 (2014).
7. A. Klug, P. Patter, K. Popovic, A. Blümel, S. Sax, M. Lenz, O. Glushko, M. J. Cordill, and E. J. W. List-Kratochvil, SPIE Org. Photonics+ Electron. **9569**, 95690N (2015).
8. H. Yan, Z. Chen, Y. Zheng, C. Newman, J. R. Quinn, F. Dötz, M. Kastler, and A. Facchetti, Nature **457**, 679 (2009).
9. H. Kipphan, *Handbook of Print Media: Technologies and Production Methods* (Springer Science & Business Media, 2001).
10. O. Glushko, M. J. Cordill, A. Klug, and E. J. W. List-Kratochvil, Microelectron. Reliab. **56**, 109 (2016).
11. O. Glushko, A. Klug, E. J. W. List-Kratochvil, and M. J. Cordill, J. Mater. Res. **32**, 1760 (2017).
12. S. Rickard, Characterization of Printed Planar Electromagnetic Coils Using Digital Extrusion and Roll-to-Roll Flexographic Processes, Clemson University, Master of Science Thesis, 2015.
13. N. Lu, X. Wang, Z. Suo, and J. J. Vlassak, Appl. Phys. Lett. **91**, 221909 (2007).
14. GOM Precise Industrial 3D Metrology, https://www.gom.com/company/contact.html (2018).
15. O. Glushko, V. M. Marx, C. Kirchlechner, I. Zizak, and M. J. Cordill, Thin Solid Films **552**, 141 (2014).
16. M. J. Cordill and A. A. Taylor, Thin Solid Films **589**, 209 (2015).
17. A. A. Taylor, M. J. Cordill, and G. Dehm, Philos. Mag. **92**, 3363 (2012).
18. M. J. Cordill, A. A. Taylor, J. Schalko, and G. Dehm, Metall. Mater. Trans. A Phys. Metall. Mater. Sci. **41**, 870 (2010).

19. V. M. Marx, M. J. Cordill, D. M. Többens, C. Kirchlechner, and G. Dehm, Thin Solid Films **624**, 34 (2017).

20. P. A. Gruber, E. Arzt, and R. Spolenak, J. Mater. Res. **24**, 1906 (2009).

21. A. A. Taylor, V. Edlmayr, M. J. Cordill, and G. Dehm, Surf. Coatings Technol. **206**, (2011).

**Figures and Captions**

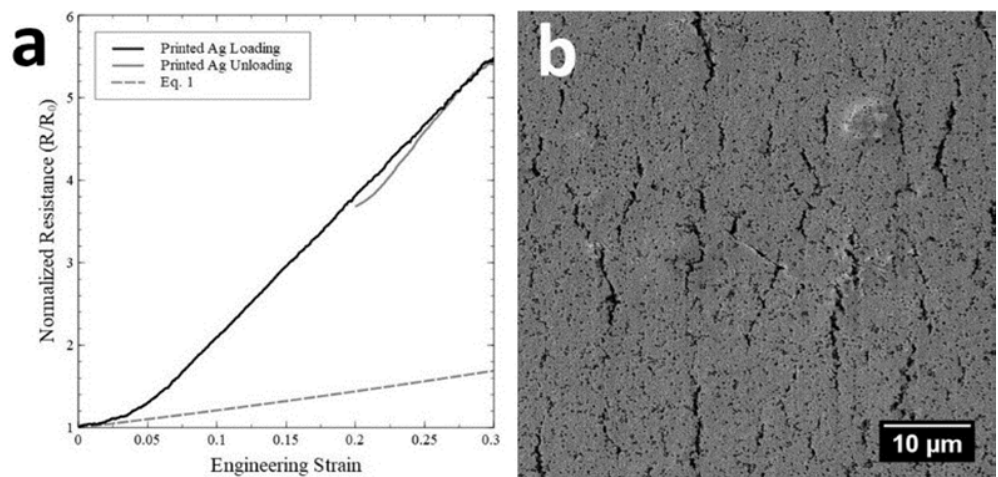

*Figure 1:* (a) Relative resistance during applied tensile load and unloading. At small tensile strains during loading, the system performed close to the predictive model until about 3% engineering strain when cracks initiated in the printed line. (b) Post-mortem SEM micrograph of the flexographically printed line after stretching to 30%. Numerous cracks that appear in black are clearly visible. The straining direction is horizontal.

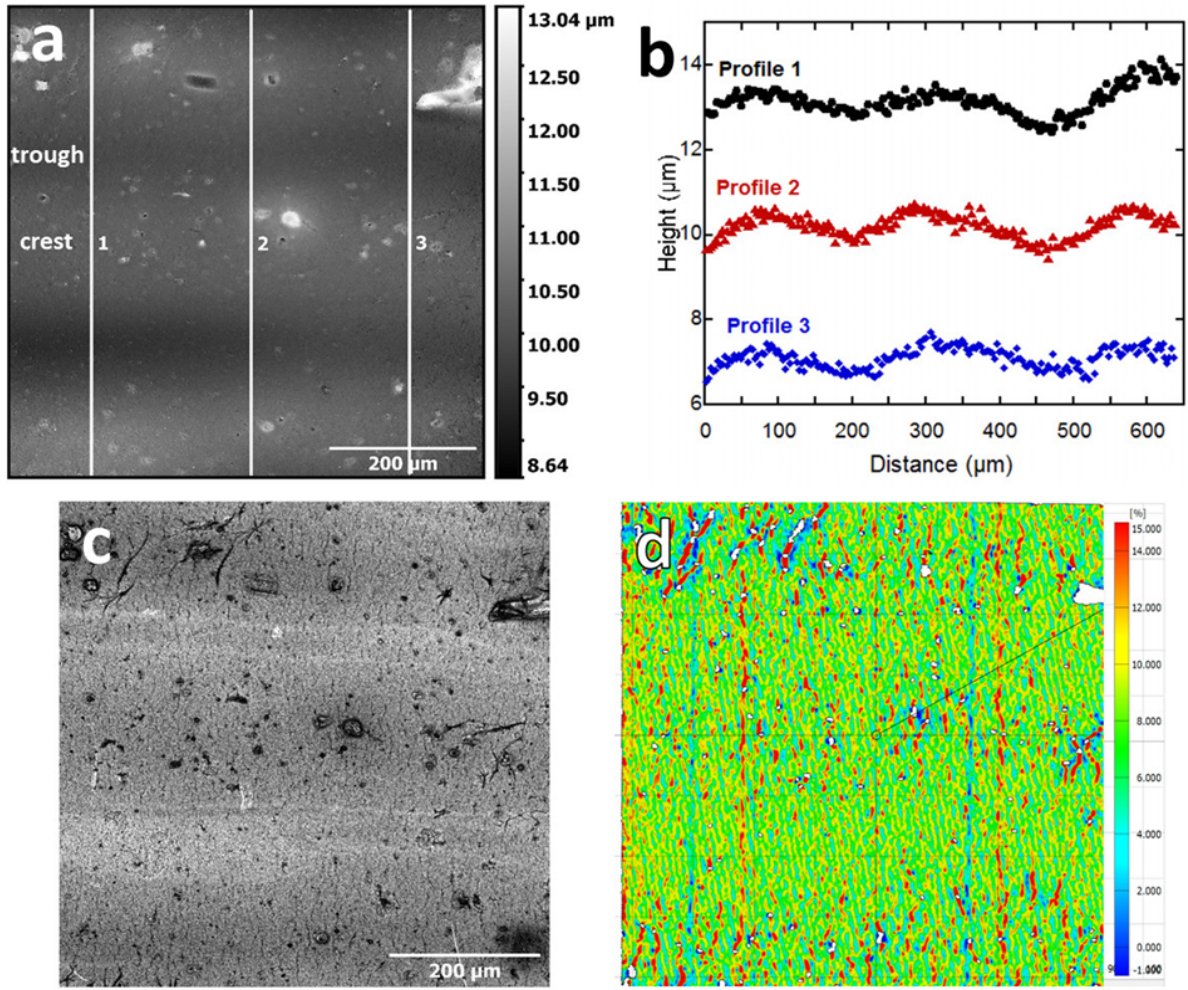

*Figure 2:* *(a) CLSM height image at 12% applied strain with three marked profiles (white lines). (b) Extracted height profiles from (a) illustrating the wavy surface of the flexographic printing process. (c) CLSM laser intensity image at 12% of the same area as (a) and (d) the DIC calculated strain image of the printed Ag line at 12% strain demonstrating that in the wave crests there are more red streaks (cracks) than in the troughs.*

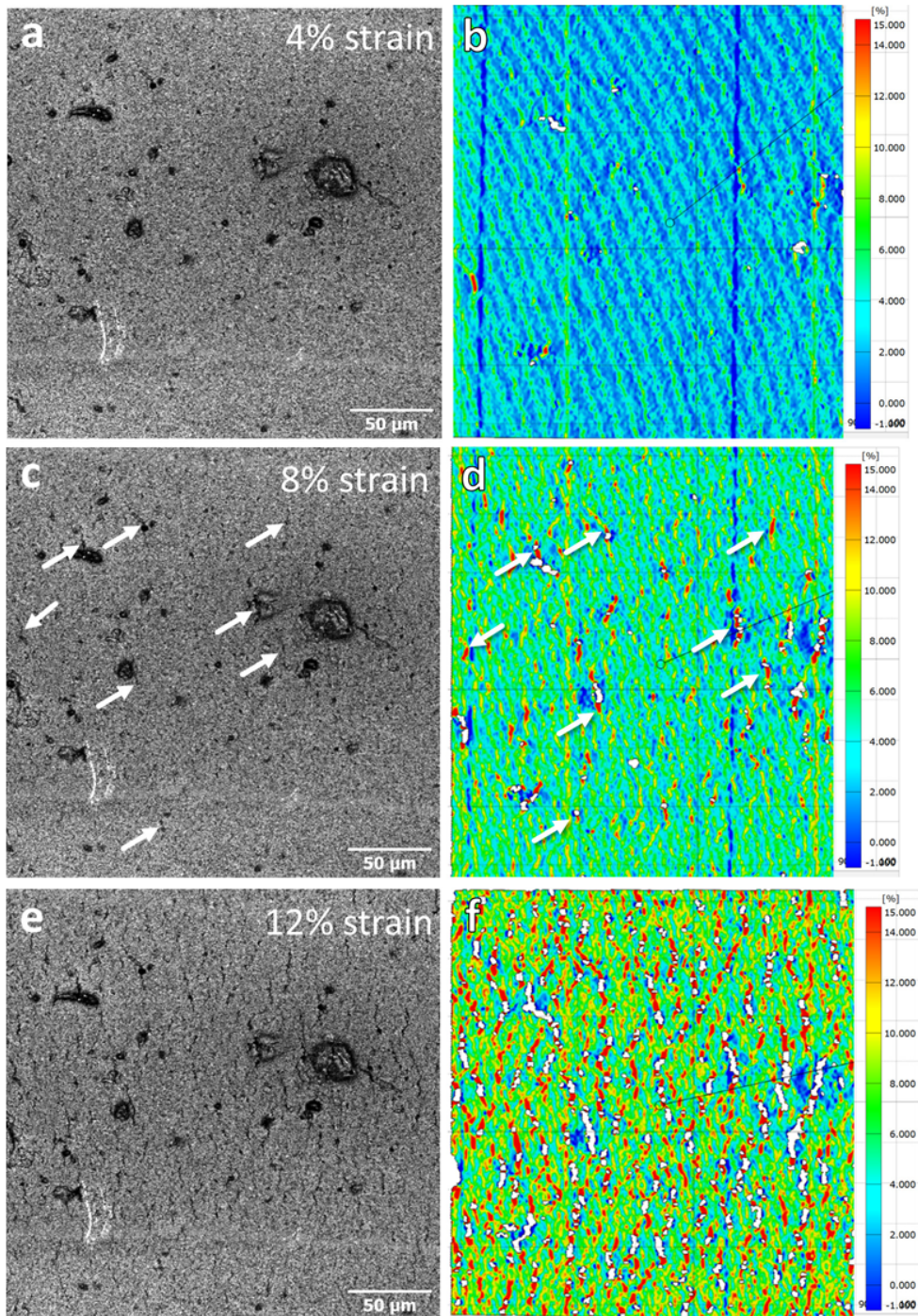

***Figure 3:*** *(a) In the CLSM laser image at 4% strain, it is difficult to observe cracks without DIC (b), which finds cracks near large defects. (c) At 8% strain cracks are observed in the CLSM laser image (arrows) and correspond well to the DIC (d). At the maximum strain of 12% cracks are more visible in the CLSM laser image (e) but the DIC evaluation (f) finds more red streaks and larger cracks (white areas).*

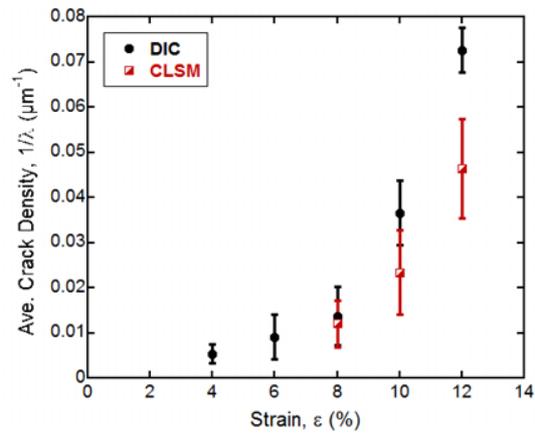

*Figure 4:* *Measured average crack density as a function of applied strain. Crack densities were evaluated from the CLSM laser images and DIC assuming areas of large tensile strain (red streaks and white areas) were cracks.*